\documentclass[prd,aps,nofootinbib]{revtex4}
\usepackage{amssymb,amsmath,epsfig}
\usepackage[bookmarks]{hyperref}
\usepackage[usenames,dvipsnames]{color}

\newcommand{\Real}{\mathbb{R}}

\begin{document}

\title{On the global structure of Kerr-de Sitter spacetimes}

\author{Kayll Lake$^{1}$, and Thomas Zannias$^{1,2}$}

\affiliation{$^1$Department of Physics, QueenÕs University, Kingston, Ontario K7L 3N6, Canada,\\
$^2$Instituto de F\'\i sica y Matem\'aticas,
Universidad Michoacana de San Nicol\'as de Hidalgo,\\
Edificio C-3, Ciudad Universitaria, 58040 Morelia, Michoac\'an, M\'exico.} 
\email{lake@astro.queensu.ca, zannias@ifm.umich.mx}

\begin{abstract}
The global structure of the family of Kerr de Sitter spacetimes is re-examined. Taking advantage of the
natural length scale set by the cosmological  constant $\Lambda>0$, conditions
on the parameters $(\Lambda, M, a^{2})$ have been found, so that a Kerr-de Sitter specetime 
either describes a black hole with well separated horizons, 
or describes degenerate configurations where two or more horizons
coincide. As long as the rotation parameter $a^{2}$
is subject to the constraint $a^{2}\Lambda  \ll 1$,
while the mass parameter   $M$ is subject to:
$ a^{2}[1+O(a^{2}\Lambda)^{2})] <M^{2}< \frac {1}{9\Lambda}[1+2a^{2}\Lambda+O(a^{2}\Lambda)^{2})]$,
then a Kerr-de Sitter spacetime with such parameters, 
describes a black hole possessing an inner horizon separated from an outer horizon
and the hole is embedded
within  a pair of cosmological horizons. Still for $a^{2}\Lambda  \ll 1$, but 
assuming that either 
$M^{2}> \frac {1}{9\Lambda}[1+2a^{2}\Lambda+O(a^{2}\Lambda)^{2})]$ or 
$M^{2}< a^{2}[1+O(a^{2}\Lambda)^{2})]$, the Kerr-de Sitter spacetime describes a ring-like singularity enclosed by two cosmological horizons. 
A Kerr-de Sitter spacetime may also describe configurations where the
inner, the outer and one of the cosmological horizons coincide. However, we found
that this coalescence
occurs provided $M^{2}\Lambda \sim 1$ and due to the observed smallness of $\Lambda$, these configurations are probably 
irrelevant in astrophysical settings. Extreme black holes, i.e. black holes where
the inner horizon coincides with the outer black hole horizon
are also admitted.
We have found that in the limit 
$M^{2}\Lambda  \ll 1$  and $a^{2}\Lambda  \ll 1$, extreme black holes
occur,  provided $a^{2}=M^{2}(1+O(\Lambda M^{2}))$.
Finally a coalescence between the outer and the cosmological horizon, although in principle possible, is likely to be 
unimportant at the astrophysical level, since this requires  $M^{2}\Lambda \sim 1$.
Our analysis shows that as far as the structure of the horizons are concerned, the family of
Kerr-de Sitter spacetimes exhibits similar structure as the  Reissner-Nordstrom-de Sitter family of
spacetimes does. 

\end{abstract}

\date{\today}


\maketitle
\section{Introduction}
 
 Since the inception of General Theory of Relativity,
 the rise, fall and the eventual reemergence of the cosmological constant $\Lambda $ has
 an interesting story\footnote{For a recent account of the history of the $\Lambda $-term
  the reader is referred to an enjoyable and well documented article by Straumann in \cite{Str1}, see also section
  $(6.6)$ in \cite{StrG1}.}. In 1917,
Einstein 
 introduced the $\Lambda $ term  into his famous equations hoping
 that the repulsive effects associated 
 with $\Lambda >0$ would lead to a static universe.
 However since observational data  favored a dynamical world model, 
  he abandoned $\Lambda $ a few years later.  With the advent of spontaneous symmetry breaking in gauge theories,
 the $\Lambda $ term re-appeared and now days  is at the epicenter  of one of the deep mysteries
 surrounding modern science. A multiple of $\Lambda $
  is interpreted as the vacuum energy density
 and the real issue is why? After so many symmetry breakings that took place
 in the early universe,  does $\Lambda $ relax to 
  the tiny value suggested by current observations? \\ 
  The recently discovered type-Ia
 cosmological supernovae provide direct observational evidences for a positive  
 cosmological constant and these developments brought $\Lambda$ back into the forefront of scientific research.
 Although current estimates suggest   $\Lambda <10^{-55} cm^{-2}$,
 nevertheless despite its tiny value $\Lambda$ has important consequences 
 on the large scale structure of the spacetime. Although it is impossible
 to summarize all the scientific work on Einsteins equations with a  
 non vanishing $\Lambda$ here, a great deal of effort has been focused on 
a family of a stationary- axisymmetric solutions 
of Einsteins equations with a non vanishing $\Lambda$, discovered long ago
 by Carter \cite{Car1},\cite{Car2}. These solutions besides $\Lambda$, contain two additional parameters $(M, a)$. 
In the limit of vanishing
 $\Lambda$, the solutions reduce to the Kerr family of metrics
  while for  $a=0$, the Schwarzschild-(anti) de Sitter
 family is recovered. Due to these properties, $M$ is interpreted as
 a mass\footnote{Whether this $M$ can be rigorously 
 interpreted as some form of mass energy
 runs into the subtleties in defining
 mass energy for asymptotically (anti) de Sitter spacetimes. For recent avances 
 consult refs. \cite{Tod}, \cite{Chr2}.}  
 and $a$ as a rotation parameter\footnote{Preliminary investigations show that
 $(M,a)$ may admit a representation in terms of scalar 
polynomial curvature invariants 
 patterning the same trend as for the case of Kerr (see the recent work in 
\cite{Lak1}). The scalar polynomial curvature invariants $(Q_{1},Q_{2})$ (see \cite{Lak1},\cite{Pag1}
 for their precise form) locate the horizons and ergosurfaces for the Kerr-(anti) de Sitter metric.
 This interesting result was not realized at the time \cite{Lak1} was written.}. For certain values of the parameters $(\Lambda, M, a)$ the solutions are interpreted as representing stationary axisymmetric black holes in asymptotically  (anti)-de Sitter background\footnote{
 For arguments supporting  this interpretation, see for instance ref.\cite{GibHaw1}. For 
 progress  on the important issue of defining a black hole on a  cosmological spacetime, see
 for instance \cite{Shi1},\cite{Shi2},\cite{Sim}.} and in contrast to black holes in an asymptotically flat spacetime, these black holes
 may  possess up to four horizons. Two of these horizons
 are  cosmological and the other two are the inner and outer black hole horizons
 enclosing a ring-like singularity.
 It is believed that this Kerr-(anti) de Sitter family, may enjoy a uniqueness property as is
 the case of the Kerr black hole and thus interpreted as the final end state of the complete gravitational  collapse of
 a bounded system in an asymptotically-(anti) de Sitter  spacetime.\\ 
  The behavior of geodesics on this space-times 
 has been the subject of many investigations, see for example \cite{Stuc1} \cite{Stuc2} \cite{Stuc3}  \cite{Poud} \cite{Zan1},
 while for the extension of the solution to arbitrary spacetime dimension see \cite{Pag2}. Effects of gravitational lensing on these 
 spacetimes has been addressed in  \cite{Kra1},\cite{Kra2},\cite{Kra3}.
 In two recent works \cite{Chr1},\cite{Mat1}, 
the global structure of this family has been addressed.
In  \cite{Chr1}, the authors introduced the notion of the 
projection diagrams as an alternative to Carter-Penrose conformal diagrams
and through these diagrams,
the structure of families of two dimensional submanifolds
 of the Kerr-(anti) de Sitter spacetime were investigated. 
In \cite{Mat1}, the authors
 advanced an interesting interpretation of the Kerr-de Sitter
spacetime, and
they raised the question regarding
the conditions upon  $(\Lambda, M, a)$ so that 
in a Kerr-de Sitter spacetime 
a coalescence between the inner and outer black hole hortizons
takes place. Via a numerical example,
 they argue that for the case where $\Lambda>0$
 the   condition $M^{2}=a^{2}$ does not any longer
 characterize an extreme Kerr-de Sitter black hole.

A complete understanding of the global structure
of the family of Kerr-(anti) de Sitter requires
an investigation of the parameter
space $(\Lambda, M, a)$. Under what restrictions upon
$(\Lambda, M, a)$ does a  Kerr-(anti) de Sitter spacetime describe
a black hole with a well  separated inner- outer and cosmological horizons?
Does there exist a non trivial subset of the parameter space where
a  Kerr-(anti) de Sitter spacetime describes extreme configurations 
i.e. configurations where for instance the cosmological horizon coincides with
the outer black horizon or do there exist  super extreme configurations
where three horizons coincide? If such configurations exist, do the  parameters $(\Lambda, M, a)$
retain values so that these configurations are important in astrophysics?
\\
 The purpose of the present work is to settle some of these questions.
 As a first step,
 we study the behavior of the roots of a quartic polynomial equation as function of the parameters 
$(\Lambda, M, a)$. We treat this problem via
the properties of the discriminate
of polynomial equations and their relations to the determinant of the Sylvester matrix.
Primarily, we focuse our attention to the parameter space  which is of relevance
for the description of astrophysical sources. Due to the currently suggested tiny
value of $\Lambda$, the restrictions $M^{2}\Lambda  \ll 1$ or (-and) $ a^{2} \Lambda  \ll 1$
cover many sources of astrophysical relevance\footnote{
Based on a value  $\Lambda \sim 10^{-55}cm^{-2}$, then 
for the case of the Sun in a  uniform rotation, 
 $M^{2}\Lambda  \sim 10^{-45}$ and $ a^{2} \Lambda  \sim 10^{-46}$
 implying that the restrictions $M^{2}\Lambda  \ll 1$ or (-and) $ a^{2} \Lambda  \ll 1$
 leave  plenty of room for the descriptions of astrophysical configurations.}.

The organization of this article is as follows. In the next section, we briefly
introduce the family of Kerr-(anti) de Sitter metrics and identify the curvature and the coordinate singularities.
In section $III$, we discus  the roots of 
a polynomial equation $\Delta (r)=0$
 and relate their occurrence to the values of the parameters  $(\Lambda, M, a)$.
In section $IV$, we comment on the global structure of the  Kerr-de Sitter spacetimes
and discusse  future work and open problems.

\section{The Kerr-de Sitter metric }
In a set of local $(t,\varphi,r,\vartheta)$ Boyer-Lindquist coordinates,
the Kerr-(anti) de Sitter family of metrics has the form   

\begin{equation}
g = -\frac {\Delta({r})}{I^{2}\rho^{2}}[dt-asin^{2}{\theta}d\varphi]^{2}+
\frac {\Delta({\vartheta})\sin^{2}\vartheta}{I^{2}\rho^{2}} [ a dt - (r^{2}+a^{2})d\varphi]^{2} 
+\frac {\rho^{2}} {\Delta(r)}d^{2}r+ \frac {\rho^{2}} {\Delta({\vartheta})}d^{2}\vartheta
\label{Eq:g}
\end{equation}
where
$$
\rho^2 := r^2 + a^2\cos^2\vartheta,\quad
\Delta (r) :=-\frac {1}{3}\Lambda r^2(r^{2}+a^{2})+r^{2}- 2M r + a^2,\quad \Delta(\vartheta) :=1+\frac {1}{3}\Lambda a^2cos^{2}\vartheta, \quad I:=1+\frac {1}{3}\Lambda a^2.
$$
For  $\Lambda>0$, this $g$ is the  Kerr-de Sitter metric while 
for $\Lambda<0$ corresponds to the Kerr-anti de Sitter metric, $(M,a^{2})$  are free parameters 
while 
the factor $I$ 
ensures the regularity
of the $g$  along the axis of axial symmetry.
The fields $\xi_{t}=\frac {\partial }{\partial t}$ and  $\xi_{\varphi}=\frac {\partial }{\partial \varphi}$
are commuting Killing vector fields with the zeros of the latter defining the rotation axis.\\
Via algebraic manipulations using $GRTensorII$ \cite{grt}, we find
\begin{equation}
C_{\mu\nu\lambda\rho}C^{\mu\nu\lambda\rho}=\frac{48M^{2}}{\rho^{12}}F(r,\vartheta), \quad F(r,\vartheta)=(r^{2}-a^{2}cos^{2}\vartheta)(\rho^{4}-16a^2r^{2}cos^{2}\vartheta),
\label{Eq:WS}
\end{equation}
\begin{equation}
C^{*}_{\mu\nu\lambda\rho}C^{\mu\nu\lambda\rho}=\frac{96M^{2}ra}{\rho^{12}}F^{*}(r,\vartheta), \quad F^{*}(r,\vartheta)=(r^{2}-3a^{2}cos^{2}\vartheta)(-3r^{2}+a^{2}cos^{2}\vartheta)cos\vartheta,
\label{Eq:WDS}
\end{equation}
where $C_{\mu\nu\lambda\rho}$ stand for the components of the Weyl, 
while 
$C^{*}_{\mu\nu\lambda\rho}$ denote the dual components.
These  invariants show that  the curvature of  (\ref{Eq:g})
becomes unbounded as $\rho \to 0 $ i.e. as the ring  ($r= 0$, $\vartheta= \frac {\pi}{2}$)
is approached. Remarkably, $\Lambda$ drops out of these invariants and  so
they exhibit the same structure as the one exhibited by the Kerr metric. More remarkably
the polynomial curvature invariants $(Q_{1},Q_{2})$ (see \cite{Lak1}) have a very simple form and
also locate the ergosurfaces and horizons.

 Coordinate singularities in
(\ref{Eq:g}) occur along 
the 
axis of axial symmetry and these singularities
can be removed by employing generalized Kerr-Schild coordinates.
The other family of coordinate singularities\footnote{For the case of Kerr-anti de Sitter $(\Lambda<0)$ an additional 
coordinate singularity may arise at the zeros of $\Delta({\vartheta})$ factor. The 
nature of this coordinate singularity, as well as more complete analysis of the Kerr-anti de Sitter spacetime,
will be discussed elsewhere \cite{LakZan1}.} in (\ref{Eq:g}),
occur  at the roots of $\Delta (r)=0$
and further ahead we discuss the  extension of (\ref{Eq:g})
through these  singularities.\\ 

Most of the analysis in the literature
has  focused on the Kerr- de Sitter metric
subject to the assumption that 
the parameters $(\Lambda, M, a)$ in  (\ref{Eq:g}),
are chosen so that
the quartic polynomial  
\begin{equation}
\Delta(r)= -\frac {1}{3}\Lambda r^{2}(r^{2}+a^{2})+r^{2}-2Mr+a^{2},\quad r\in \Real
\label{Eq:D}
\end{equation}
has one  negative zero and the three distinct  positive ones.
Although for this case the Kerr-de Sitter family of spacetimes
exhibits rich structure, nevertheless this family 
contains other configurations as well. A complete classification of all possible configurations,
requires an understanding of the roots of the quartic $\Delta (r)=0$
as a function of the parameters $(\Lambda, M, a)$ and in the next section,
we discuss that problem.

\section{On the roots of the equation $\Delta(r)=0$  }
It  is convenient
for the purposes of this section, to introduce a set of abbreviations so that 
 \begin{equation}
\Delta (r)= -\frac {1}{3}\Lambda r^{4} +(1-\frac {1}{3}\Lambda a^{2})r^{2}-2Mr+a^{2}:=p_{4}r^{4}+p_{3}r^{3}+p_{2}r^{2}+p_{1}r+p_{0},\quad r \in \Real
\label{Eq:D1}
\end{equation}
\begin{equation}
p_{4}= -\frac {1}{3}\Lambda:=L, \quad p_{3}=0,\quad p_{2}=1-\frac {1}{3}\Lambda a^{2}:=N, \quad p_{1}=-2M:=K,\quad p_{0}=a^{2}.
\label{Eq:DC}
\end{equation}
Furthermore, here after, $\Delta^{'}(r)$, $\Delta^{''}(r)$  stand for the first and second derivatives
of the polynomial $\Delta(r)$ and $r_{i}, i \in (1,2,3,4)$ denote the roots of  $\Delta (r)=0$.\\
The discriminant $D(\Delta_{r})$
of the polynomial equation $\Delta (r)=0$
is defined by 
\begin{equation}
D(\Delta_{r})=p_{4}^{6}\prod_{\substack{i<j}}(r_{i}-r_{j})^{2},\quad i,j \in (1,2,3,4)
\label{Eq:DIC}
\end{equation}
and satisfies the important relation:
\begin{equation}
D(\Delta_{r})=\frac {R(\Delta,\Delta^{'})}{p_{4}} 
\label{Eq:DIS}
\end{equation}
where $R(\Delta,\Delta^{'})$ is  the determinant of the Sylvester matrix 
associated with the polynomials  $\Delta(r)$ and  $\Delta^{'}(r)$. The determinant of the Sylvester 
matrix can be computed in terms of the coefficients of 
$\Delta(r)$ and $\Delta^{'}(r)$ and thus (\ref{Eq:DIC}, \ref{Eq:DIS})
provide insights regarding the reality and multiplicity 
of the roots of $\Delta (r)=0$. (For an introduction to the theory
leading to the fundamental identity (\ref{Eq:DIS}) see for instance \cite{Ede},\cite{Pra}.)
Although the 
evaluation of the Sylvester determinant  can be a tedious job,
 fortunately for polynomials of low order, it has been tabulated and the result are readily available in the literature.
For the polynomial $\Delta(r)$, consulting Maple, Mathematica, or \cite{Ede},\cite{Pra}, we find
\begin{equation}
D(\Delta_{r})=256a^{6}L^{3}-128a^{4}N^{2}L^{2}+144a^{2}K^{2}NL^{2}+16a^{2}N^{4}L-27K^{4}L^{2}-4K^{2}N^{3}L\nonumber
\end{equation}
\begin{equation}
=128[2a^{6}L^{3}-a^{4}N^{2}L^{2}+\frac {9}{8}a^{2}K^{2}NL^{2}+\frac {1}{8}a^{2}N^{4}L-\frac {27}{128}K^{4}L^{2}-
\frac {1}{32}K^{2}N^{3}L]
\label{Eq:DISS}
\end{equation}
while the discriminants for $\Delta^{'}(r)$ and $\Delta^{''}(r)$ have the form
\begin{equation}
D(\Delta^{'}_{r})=-16[27K^{2}L^{2}+8N^{3}L],\quad D(\Delta^{''}_{r})=-96LN.
\label{Eq:DISD}
\end{equation}

Hereafter we consider only the cases where  $\Lambda> 0$ and  $a\neq 0$. The analysis of Kerr-anti de Sitter
is discussed elsewhere \cite{LakZan1}.
For $\Lambda> 0$ and  $a\neq 0$, the equation $\Delta(r)=0$ has 
 at least one negative and one positive root.
 The widely discussed case of Kerr-de Sitter metric,
 assumes that the equation
$\Delta(r)=0$
admits one negative  and three distinct positive roots and this occurs provided
\begin{equation}
 D(\Delta_{r})>0,\quad \Lambda>0,\quad a\neq 0.
\label{Eq:FDR}
\end{equation}
The condition $D(\Delta_{r})>0$ by itself 
implies that either all the roots of $\Delta (r)=0$ are real and distinct or 
they form two pairs of complex conjugate roots. However this last possibility
is eliminated once the  three conditions in (\ref{Eq:FDR})
are taken together.

To get insights into the nature of the restrictions that conditions (\ref{Eq:FDR}) impose
upon $(\Lambda, M, a^{2})$, at first we write the discriminant $D(\Delta_{r})$ in $(\ref{Eq:DISS})$  in the equivalent form
\begin{equation}
 D(\Delta_{r})=-\frac {128}{3}[AM^{4}+BM^{2}+C],\quad A=\frac {9}{8}\Lambda^{2},\quad B=-\frac {N\Lambda}{8}(
 N^{2}+12a^{2}\Lambda),\quad C= [\frac {2}{9}a^{4}\Lambda^{2}+\frac {1}{3}a^{2} \Lambda N^{2}
  +\frac {N^{4}}{8}] a^{2}\Lambda, 
 \label{Eq:DISSE}
\end{equation}
while a computation shows 
\begin{equation}
 B^{2}-4AC=\frac {\Lambda^{2}}{64}[N^{6} -
 12a^{2}\Lambda N^{4} +48a^{4}\Lambda^{2}N^{2}-64a^{6}\Lambda^{3}]\equiv\frac {\Lambda^{2}}{64}T(a^{2}\Lambda)
    \label{Eq:DISDI}
\end{equation}
where  $T(a^{2}\Lambda)$ is a sixth order polynomial with respect to the positive variable $a^{2}\Lambda$.
The graph 
of this polynomial, determines domains on the $a^{2}\Lambda$-axis, where  it
is positive definite, negative definite or zero.
For any 
$a^{2}\Lambda$ within the domains where $T(a^{2}\Lambda)$
is positive definite, $B^{2}-4AC$ is positive definite and thus   $AM^{4}+BM^{2}+C=0$ 
has real positive roots  $\rho_{-}(\Lambda,a^{2})< \rho_{+}(\Lambda,a^{2})$. 
For any $M^{2}$ subject to the bounds 
\begin{equation}
\rho_{-}(\Lambda,a^{2})< M^{2}<\rho_{+}(\Lambda,a^{2})
  \label{Eq:COMP}
 \end{equation}
 the inequality $D(\Delta(r))>0$ holds. 
 However since the observational data suggest  
a tiny value for  $\Lambda$, it is reasonable to focus our attention
to the case where
$a^{2}$ and $\Lambda$ are chosen so that $ a^{2} \Lambda  \ll 1$.
Although this is a strong restriction,
neverthesess it is satisfactory from the
astrophysical view point since it 
 covers wide range of  astrophysical sources.
Assuming therefore that  $ a^{2} \Lambda  \ll 1$, 
the roots $\rho_{-}(\Lambda,a^{2})< \rho_{+}(\Lambda,a^{2})$ of $AM^{4}+BM^{2}+C=0$
are 
\begin{equation}
\rho_{+}(\Lambda,a^{2})=\frac {1}{9\Lambda}[1+2a^{2}\Lambda+O(a^{2}\Lambda)^{2})],\quad
\rho_{-}(\Lambda,a^{2})=a^{2}[1+O(a^{2}\Lambda)^{2}] 
   \label{Eq:ROOTS}
\end{equation}
implying that $D(\Delta(r))>0$, provided 
$M^{2}$ lies in the domain:
\begin{equation}
 a^{2}[1+O(a^{2}\Lambda)^{2}] <M^{2}< \frac {1}{9\Lambda}[1+2a^{2}\Lambda+O(a^{2}\Lambda)^{2})].
   \label{Eq:PQ}
\end{equation}
This condition as far as we are aware, is new. It is fundamental and asserts that 
as long as $M^{2}$ is chosen to satisfy these bounds
and $a^{2}\Lambda  \ll 1$, then $\Delta (r)=0$, admits three real positive distinct roots and a negative
 one. This estimate gives a relation between the mass $M$, rotation parameter $a^{2}$ and  $\Lambda$ so that
 a Kerr-de Sitter spacetime describes a black hole possessing an inner, outer and two cosmological horizons.

A modification of the conditions in (\ref{Eq:FDR}) covers the case where $\Delta (r)=0$ 
admits one negative, one positive and
a pair of complex conjugate roots.
This can occur, provided
\begin{equation}
 D(\Delta_{r})<0,\quad \Lambda>0,\quad a\neq 0.
\label{Eq:FDRC}
\end{equation}
Based on the same reasoning as above,
we assume  $ a^{2}\Lambda  \ll 1$ and thus  the condition $D(\Delta_{r})<0$ holds, provided either
$M^{2}> \frac {1}{9\Lambda}$ or  $M^{2}<  a^{2}$. A Kerr-de Sitter metric with parameters in that range
describe a ring like  curvature singularity enclosed between a pair of cosmological horizons.
It is interesting to note that the parameter space is dominated by regions  where the equation
$\Delta (r)=0$ has only a pair of real roots and this property has some interesting ramifications regarding
the validity of cosmic censorship  within the cosmological domain.
However it is important  to stress that the dominance of the 
parameter space by regions where
$\Delta (r)=0$ possess a  pair of real roots
holds under validity of the restriction 
$a^{2}\Lambda  \ll 1$. Dropping this restriction
likely will alter this conclusion. 
Although it is interesting to analyze the case where  the condition $a^{2}\Lambda  \ll 1$ is relaxed, 
we shall not proceed with this case any further here (see however, comments further ahead).\\
For completeness, we now investigate the case where $\Delta(r)=0$ admits multiple roots
and as a first case we treat the case where $\Delta(r)=0$ admits
a negative root and a positive root of multiplicity
three (or the closely related alternative of a negative root of multiplicity three and a simple positive root).
From the properties of the discriminant, it is easily seen that this  setting  occurs provided:
\begin{equation}
D(\Delta_{r})=D(\Delta{'}_{r})=0,\quad D(\Delta{''}_{r})>0, \quad
\Lambda>0,\quad a\neq 0.
\label{Eq:NR}
\end{equation}
The conditions $D(\Delta_{r})=D(\Delta{'}_{r})=0$ guarantee that 
$\Delta(r)=0$ has a real root $r_{i}$ of multiplicity at least three,
while  $D(\Delta{''}_{r})>0$ implies that
$r_{i}$ has multiplicity three.

Since $D(\Delta^{'}_{r})=-16L[27K^{2}L+8N^{3}]=\frac {64\Lambda}{3}[-9M^{2}\Lambda+2N^{3}]$ and $\Lambda>0$, 
clearly $D(\Delta^{'}_{r})=0$
cannot be  satisfied unless $N>0$. In turn,  $N>0$  requires  $a^{2}\Lambda<3$
and under validity of this constraint\footnote{In the alternative case, i.e. whenever  $a^{2}\Lambda\geq3$ and 
$\Lambda>0$, the disciminant  $D(\Delta^{'}_{r})$ is always negative definite
which implies further that $\Delta(r)=0$ has only a pair of real roots.}, $D(\Delta^{'}_{r})=0$ demands
\begin{equation}
M^{2}=\frac {2N^{3}}{9\Lambda}=\frac {2}{9\Lambda} (1-\frac {1}{3}a^{2}\Lambda)^{3}.
\label{Eq:NR1}
\end{equation}
Since $\Delta{''}(r_{i})=0$ has  $r_{\pm}=\pm(\frac {N}{2\Lambda})^
{\frac {1}{2}}$ as its roots, there exist
two possibilities regarding
the triple root of $\Delta(r)=0$.
Choosing $r_{i}:=r_{+}=(\frac {N}{2\Lambda})^
{\frac {1}{2}}>0$ then $\Delta{'}(r_{+})=0$ provided the positive root is taken 
in (\ref{Eq:NR1}) ie
\begin{equation}
M_{+}=\frac {{2}^{\frac {1}{2}}}{3} \frac {N^{\frac {3}{2}}}{{\Lambda}^{\frac {1}{2}}},
\label{Eq:NRP}
\end{equation}
while  $r_{i}:=r_{-}=-(\frac {N}{2\Lambda})^
{\frac {1}{2}}<0$ obeys  $\Delta{'}(r_{-})=0$ provided
\begin{equation}
M_{-}=-\frac {{2}^{\frac {1}{2}}}{3} \frac {N^{\frac {3}{2}}}{{\Lambda}^{\frac {1}{2}}}.
\label{Eq:NRN}
\end{equation}
Finally $r_{+}$  (respectively $r_{-}$) is also a root of $\Delta(r)=0$, provided 
\begin{equation}
a^{2}=\frac {{N}^{2}}{4\Lambda}=\frac {1}{4\Lambda}(1-\frac {1}{3}a^{2}\Lambda )^{2}.
\label{Eq:A}
\end{equation}
Setting $y=a^{2}\Lambda$, this constraint yields  to the quadratic equation $y^{2}-42y+9=0$
with roots $2y=42\pm\sqrt{1728}\simeq 42\pm41.56$
and thus  for any choice of $\Lambda>0$, exist a 
 value for $a^{2}\Lambda$ consistent with the constraint
$a^{2}\Lambda <3$. Moreover for  the values $M^{2}$ and $a^{2}$
as in (\ref{Eq:NR1}, \ref{Eq:A}), it can be seen that the discriminant $D(\Delta_{r})$ vanishes
identically.
In summary, the Kerr de Sitter family
allows configurations where the inner, outer and cosmological horizon coincide.
The location of this triple horizon and the required (positive mass) are\footnote{
For any choice of $\Lambda >0$, equation (\ref{Eq:A}) determines a value of
$a^{2}\Lambda$ and thus for this value of $a^{2}\Lambda$, 
(\ref{Eq:COOR}) determine
the location of the triple root and enclosed mass.} :
\begin{equation}
r_{+}=(\frac {N}{2\Lambda})^
{\frac {1}{2}},\quad
M_{+}=\frac {{2}^{\frac {1}{2}}}{3} \frac {N^{\frac {3}{2}}}{{\Lambda}^{\frac {1}{2}}}.
\label{Eq:COOR}
\end{equation}
Due to the tiny value of the observed $\Lambda$ and since $N\simeq 1$, the occurrence of a triple root 
requires extremely high values of the mass parameters $M^{2}$
and likely these configurations are irrelevant for the description of astrophysical  systems.\\

We finish this section by considering the case 
where $\Delta(r)=0$ has real roots but one of them has multiplicity two\footnote{The possibility
that there exist two roots both of multiplicity two
it is not compatible  with $\Lambda>0$ and $a^{2}\neq 0$.}.
This arrangement can occur in one of the forms;
\begin{equation}
r_{1}=r_{2}< r_{3}<r_{4},\quad r_{1} <r_{2}=r_{3}<r_{4},\quad r_{1}<r_{2}<r_{3}=r_{4}.
\label{Eq:DROOTS}
\end{equation}
Again, in view of the properties of the discriminant, this setting occurs provided
\begin{equation}
 D(\Delta_{r})=0,\quad D(\Delta{'}_{r})>0, \quad \Delta(\hat r_{i})=0, \quad \Delta^{''}(\hat r_{i})\neq0 
 \label{Eq:FC}
\end{equation}
where $\hat r_{i}$ stands for any of the roots of $\Delta {'}(r)=0$ (assuming for the moment all of them real and distinct).
The condition $D(\Delta_{r})=0$ guarantees  that $\Delta(r)=0$ admits (at least one) multiple root,
$D(\Delta{'}_{r})>0$ guarantees that $\Delta {'}(r)=0$ has three real
and distinct roots , while 
 $\Delta(\hat r_{i})=0$ combined with $\Delta^{''}(\hat r_{i})\neq0$
 implies that $\hat r_{i}$ is just a double root of $\Delta(r)=0$. 

If $ \Delta(r)=0$ and $\Delta{'}(r)=0$ share a common root denoted  by 
$\hat R$,
then necessarily $(M,a^{2})$ are related to this root via   
\begin{equation}
 M={\hat R}(N-\frac {2\Lambda}{3}{\hat R}^{2}),\quad a^{2}={\hat R}^{2}(N-\Lambda {\hat R}^{2}).
 \label{Eq:Ma}
\end{equation}
The first relation $M(\Lambda, \hat R)$ is
just a restatement that $\hat R$ is a root of  $\Delta{'}(r)=0$ while 
$a^{2}(\Lambda, \hat R)$,  is the necessary and sufficient condition that 
$\hat R$ is a root of $ \Delta(r)=0$ given that $\hat R$ is a root of 
 $\Delta{'}(r)=0$. For these choices, $ \Delta(r)$ has a double zero
 and thus 
$D(\Delta_{r})=0$.\\
For  $\Lambda> 0$, the condition
$D(\Delta{'}_{r})>0$ requires
\begin{equation}
a^{2}\Lambda<3, \quad M^{2}< \frac {2}{9\Lambda}(1-\frac{1}{3}a^{2}\Lambda)= \frac {2}{9}\frac {N^{3}}{\Lambda}
\label{Eq:DR}
\end{equation}
and under these  restrictions, $\Delta^{'}(r)=-\frac {4\Lambda}{3}[r^{3}+c_{1}r+c_{0}]:=-\frac {4\Lambda}{3}c(r)=0,~ c_{1}=-
\frac {3N}{2\Lambda},~c_{0}=\frac {3M}{2\lambda}$, possess three real roots\footnote{
If  $r_{0}$ is root of $c(r)=0$, then via Cardano's method 
we set  $r_{0}=u+v$ and introduce $(\alpha, \gamma)$ so that  $\alpha=u^{3}, \gamma=v^{3}$.
In this representation $r_{0}$
is a root of $c(r)=0$, provided $\alpha\gamma=(-\frac {c_{1}}{3})^{3}$
and  $\alpha+\gamma+c_{0}=0$ and thus  $(\alpha, \gamma)$ 
are the roots of $x^{2}+c_{0}x+(-\frac {c_{1}}{3})^{3}=0$, $x \in \Real $. If $\hat \Delta$ is the discriminant 
of this equation, then the requirement $D(\Delta{'}_{r})>0$ implies $\hat \Delta<0$ and thus  $2\alpha=-c_{0}+i \sqrt {\arrowvert {\hat \Delta} \arrowvert }$ while  $\gamma$ is the complex conjugate of $\alpha$. In polar representation, $\alpha=\hat {\rho}e^{i\hat \vartheta}$ where $\hat \rho$ and $\hat \vartheta$
are as in (\ref{Eq:MR}) with the angle $\hat \vartheta$  measured counterclockwise from the positive real axis.
The three real roots of $\Delta^{'}(r):=-\frac {4\Lambda}{3}c(r)=0$ are then the
 three distinct fractional powers:
 $\alpha^{\frac {1}{3}}+\gamma^{\frac {1}{3}}$ which in polar represenations are as in (\ref{Eq:RD}).} 
\begin{equation}
\hat r_{1}=2\hat {\rho}^{\frac {1}{3}}cos(\frac {{\hat \vartheta}}{3}),\quad \hat r_{2}=2\hat {\rho}^{\frac {1}{3}}
cos(\frac {{\hat \vartheta}}{3}+\frac {2\pi}{3}),\quad \hat r_{3}=2\hat {\rho}^{\frac {1}{3}}
cos(\frac {{\hat \vartheta}}{3}+\frac {4\pi}{3})
\label{Eq:RD}
\end{equation}
with  $\hat \rho^{2}$ and the phase angle ${\hat \vartheta}$ given by:
\begin{equation}
\hat \rho^{2}=\frac {N^{3}}{8\Lambda^{3}}=\frac {1}{8{\Lambda}^{3}}(1-\frac {1}{3}\Lambda a^{2})^{3},\quad  
cos{\hat \vartheta}=-\frac {M}{x},\quad x^{2}=\frac {2}{9}\frac {N^{3}}{\Lambda}.
\label{Eq:MR}
\end{equation}
 In order that any of the roots $\hat r_{i}$ in  (\ref{Eq:RD}) 
is  simultaneously  a root of $\Delta(r)=0$, requires that  the  value of $a^{2}$ resulting from  (\ref{Eq:Ma})
once $\hat R$ is substituted for the chosen $\hat r_{i}$,
to be  positive definite and moreover be compatible with
the constraints in (\ref{Eq:DR}). 
In order to get insights into the conditions leading to the appearance of
double roots, we treat the case where 
$M^{2}\Lambda  \ll 1$
and $a^{2}\Lambda  \ll 1$. In that regime, 
the phase angle  ${\hat \vartheta}$ in (\ref{Eq:MR})
can be approximated by
\begin{equation}
{\hat \vartheta}=\frac {\pi}{2}+\frac {M}{x} +O(\frac {M}{x})^{2},\quad x^{2}=\frac {2}{9}\frac {N^{3}}{\Lambda}
\label{Eq:APA}
\end{equation}
and if we assume $M>0$, the roots in  (\ref{Eq:RD}) 
can be approximated by:
\begin{equation}
\hat r_{1}=\sqrt {\frac {3}{2}} \frac {1}{\sqrt{\Lambda}}[1-\frac {M\sqrt {\Lambda}}{\sqrt {6}}+O(\frac {M}{x})^{2}],\quad
\hat r_{2}=-\sqrt {\frac {3}{2}} \frac {1}{\sqrt{\Lambda}}[1+\frac {M\sqrt {\Lambda}}{\sqrt {6}}+O(\frac {M}{x})^{2}]
,\quad \hat r_{3}=M(1+O(\frac {M}{x})^{2}).
\label{Eq:AROOTS}
\end{equation}
Thus two roots are positive and one is negative, an expected conclusion 
based on the structure of the equation $\Delta^{'}(r)=0$  for positive $M$. 
Upon substituting $\hat r_{1}$ or $\hat r_{2}$
into the right hand side  (\ref{Eq:Ma}), we obtain a negative value for $a^{2}$
and thus $\hat r_{1}$ or $\hat r_{2}$ cannot be the location of the double root.
However, the choice $\hat r_{3}$ gives $a^{2}=M^{2}(1+O(\Lambda M^{2}))$ which is compatible with 
the constraints\footnote{The analysis for $M<0$ case yields similar results except that now 
in (\ref{Eq:AROOTS}) two of
the  roots are negative and one is positive. The interpretation of these results are of course identical to those
in case of $M>0$.}
 in (\ref{Eq:DR}). Thus in the regime $M^{2}\Lambda  \ll 1$
and $a^{2}\Lambda  \ll 1$, there is the possibility of the occurrence of a double root
at the value $\hat r_{3}\simeq M$ provided $a^{2}=M^{2}(1+O(\Lambda M^{2}))$.\\

In order to complete the picture regarding the formation of double roots, we examine
the  case where  $M$ approaches the limiting value  $x$ from bellow. Recalling that $x^{2}$ is defined in (\ref{Eq:APA}), and setting $M=x(1-\epsilon)$  with $0<\epsilon  \ll 1$
 then  (\ref{Eq:RD}) in this regime yields the approximated roots
 
 \begin{equation}
{\hat \vartheta}=\pi-\sqrt {2\epsilon},\quad \epsilon>0
\label{Eq:APAO}
\end{equation}
\begin{equation}
\hat r_{1}=\frac {1}{\sqrt {2\Lambda}}[1- \sqrt {\frac {2\epsilon}{3}}+O(\epsilon)],\quad
\hat r_{2}=\frac {2}{\sqrt {2\Lambda}}[-1+O(\epsilon)],\quad
\hat r_{3}=\frac {1}{\sqrt {2\Lambda}}[1+ \sqrt {\frac {2\epsilon}{3}}+O(\epsilon)].
\label{Eq:AROOTSO}
\end{equation}
 However, $\hat r_{2}$ cannot be a double of $\Delta(r)=0$,
since the resulting $a^{2}$ turns out to be negative. For the other two roots 
we get  $a^{2}=(4\Lambda)^{-1}(1+O(\epsilon))$
which suggests within our approximation,
$(\hat r_{1}, \hat r_{3})$ could be the location of a double root for $\Delta(r)=0$. 
In summary therefore and for values of $M^{2}$ approaching
the scale $\Lambda^{-1}$ from bellow, there is 
 the possibility of the occurrence of a double root in $\Delta(r)=0$ 
 at cosmological length scales.\\
We conclude this section by comparing
 the results derived so far 
 with known results valid for the Reissner-Nordstrom-de Sitter family
 of  spacetimes. This family, originally discovered  by Kotller  \cite{Kot} but also
  appear as a special case of Carter's family of metrics 
  derived  in \cite {Car1},\cite {Car2}. It has been widely discussed in the literature
  and for properties and  references, see for instance \cite {Rom},\cite {Bril},\cite {Lak2}.
  In a suitable set of spherical coordinates,  the global structure of this family is determined by 
    the function $\hat \Delta(r)$ defined by   
\begin{equation}
F(r)=1-\frac {2M}{r}+\frac {Q^{2}}{r^{2}}-\frac {\Lambda}{3}r^{2}=\frac {\hat \Delta(r)}{r^{2}},\quad  {\hat \Delta(r)}=:-\frac {\Lambda}{3} r^{4} +r^{2}-2Mr+Q^{2}.
\label{Eq:RND}
\end{equation}
 This $\hat \Delta(r)$ can be obtained from eq. (\ref{Eq:D1})
by taking $N(r):=1$
and replacing  $a^{2}$ by $Q^{2}$ with the latter interpreted as  the electric charge
in the solution. Therefore 
 the results of this section are also applicable 
for the 
Reissner-Nordstrom-de Sitter family of spacetimes.\\

Assuming $\Lambda>0$ and $Q^{2}\neq 0$,
it is seen from 
 eqs. (\ref{Eq:NR1}) and (\ref{Eq:A}) 
 that $\hat \Delta(r)=0$
 admits a triple positive  root
 provided $(\Lambda, M, Q^{2})$ obey the conditions:
 $9M^{2}\Lambda=2$ and $4\Lambda Q^{2}=1$. These conditions agree
  with those obtained in \cite{Rom}, \cite{Bril} and in  the terminology of \cite{Bril},
  this case is referred as the '' ultra extreme "  Reissner-Nordstrom-de Sitter spacetime.
   Setting $N(r)=1$ and replacing $a^{2}$ by $Q^{2}$ in
   (\ref{Eq:DISDI}), we find that in the limit $Q^{2}\Lambda \ll 1$, that the eq. $\hat \Delta(r)=0$ has three distinct 
   positive real roots
   provided 
  \begin{equation}
 Q^{2}[1+O(Q^{2}\Lambda)^{2})] <M^{2}< \frac {1}{9\Lambda}[1+3Q^{2}\Lambda+O(Q^{2}\Lambda)^{2})]
   \label{Eq:RNL}
\end{equation}
which to the required order agrees with the results in \cite{Bril}. For $M^{2}$ away from these domain, but still
within the regime $Q^{2}\Lambda \ll 1$,
a Reissner-Nordstrom-de Sitter spacetime admits only a cosmological horizon, referred in  \cite{Bril}
as the generic naked singularity case. For particular values $(\Lambda, M, Q^{2})$, the equation 
$\hat \Delta(r)=0$ admits two  positive roots with the one having multiplicity two. These configurations
describe extreme Reissner-Nordstrom-de Sitter spacetimes where either the inner and outer black hole coincide
or the outer horizon coincides with the cosmological horizon. Under the conditions, 
 $M^{2}\Lambda  \ll 1$ and  $Q^{2}\Lambda \ll 1$ our results show
 that the first possibility occurs under the condition  $Q^{2}=M^{2}(1+O(M^{2}\Lambda ))$, while
 the second possibility requires $M^{2}\Lambda \sim 1$.\\
 
After the completion of this work, we become aware
of a thesis \cite{Olz} written by one of the authors in  \cite{Chr1},
where a detailed analysis of the roots of the eq. $\Delta(r)=0$
 is presented.
The starting point in \cite{Olz}, is the quartic polynomial 
  \begin{equation}
 P(x)=-x^{4}+3x^{2}-2\beta x+\gamma,\quad  r=\sigma x,\quad \beta=\frac {3M}{\Lambda \sigma^{3}}, \quad \gamma=\frac {3a^{2}}{\Lambda \sigma ^{4}},\quad \sigma=(\frac {N}{\Lambda})^{\frac {1}{2}},
  \label{Eq:CHR}
\end{equation}
which is equivalent to the polynomial 
$\Delta(r)$ in (\ref{Eq:D1}). 
The formulas (\ref{Eq:DISS}) and (\ref{Eq:DISD}),
imply that the discriminants 
of  $P(x)$ and its derivative $P'(x)$ are
\begin{equation}
 D(P_{x})=-16.27[\beta^{4}-(4\gamma +1)\beta^{2}+\frac {16}{27}\gamma^{3}+\frac {8}{3}\gamma^{2}+3\gamma],\quad
 D(P'_{x})=-27.48[\beta^{2}-2] 
  \label{Eq:DISP}
\end{equation}
and in   [27], an analysis of the roots $\beta^{2}_{-}(\gamma), \beta^{2}_{+}(\gamma)$
of the equation  $D(P_{x})=0$ has been made.
It  is shown that as long as $M> 0$,
and $\gamma \in [0,\frac {3}{4}]$, then $D(P_{x})>0$
provided
$\beta^{2}_{-}(\gamma)< \beta^{2}<\beta^{2}_{+}(\gamma)$
and this condition is the analogue of our eq. (\ref{Eq:COMP}).
The condition that  $D(P'_{x})=0$
requires $\beta^{2}=2$ which is identical to our eq. (\ref{Eq:NR1}) which resulted 
upon imposing $D(\Delta{'}_{r})=0$. We have checked that the conclusions reached in \cite{Olz}, are in accord with the results
obtained in this section (within the approximation employed in this work). 
The advantage of the approach in \cite{Olz}
lies in the simple form of the polynomial $P(x)$ that allowed an analytical treatment of the 
roots of the equation $D(P_{x})=0$ in terms of the parameters $(\beta, \gamma)$.
The latter are however, complicated expressions 
of the parameters $(\Lambda, M, a^{2})$. In contrast, in this work we 
strived to obtain conditions upon $(\Lambda, M, a^{2})$
so that a Kerr-de Sitter spacetime could be employed to model 
astrophysical sources, naturally therefore our analysis has been restricted
 to a limiter region of the parameter space.\\
 Finally in \cite{Stuc3}, by a combination of analytical and numerical methods,
 conditions upon $(\Lambda, M, a^{2})$ have found
 so that a Kerr-de Sitter spacetime describes a black hole embedded
within two cosmological horizons. Although qualitatively the results  in \cite{Stuc1}
agree with those obtained here, due to different
methods and approximations no further comparison can be made.

\section{Discussion}
In this work, we have re-examined the Kerr-de Sitter family of spacetimes and the
results add complimentary insights on this structurally rich family
of spacetimes. The conclusion that whenever $a^{2}\Lambda  \ll 1$
and $ a^{2}[1+O(a^{2}\Lambda)^{2})] <M^{2}< \frac {1}{9\Lambda}[1+2a^{2}\Lambda+O(a^{2}\Lambda)^{2})]$,
 then a Kerr-de Sitter metric describes
a black hole within pair of cosmological horizons,
illustrates the role of a positive cosmological constant 
upon the black hole structure. When $M^{2}$ approaches $a^{2}$ from above,
the inner and outer black hole horizon tend to coalesce, while at the othe extreme i.e. as 
$M^{2}$ approaches the limiting lenght  scale $\Lambda^{-1}$ from bellow, 
the outer horizon tends to coalesce with the cosmological horizon.
 These  conclusions show that a non vanishing positive cosmological constant sets limit on the black hole size
in accord with results obtained in  \cite{Shi1},\cite{Shi2},\cite{Sim}.\\ 
Even though our results juxtapose  the Kerr-de Sitter family of spacetime 
with the familiar Kerr family, in addition they offer further insights on the global structure 
of these spacetimes. Starting from  
a local Boyer-Lindquist $(t,\varphi,r,\vartheta)$ set of coordinates with
$r_{i}<r< r_{i+1}$ where $r_{i}, r_{i+1}$
are two consecutive zeros
of $\Delta (r)$, then in a set of ingoing Finkelstein coordinates
$(v, \overleftarrow{\varphi},r,\vartheta)$ defined by 
\begin{equation}
dv=dt +\frac {I(r^{2}+a^{2})}{{\Delta_{r}}}dr,\quad d\overleftarrow{\varphi}=d\varphi+\frac {Ia}{{\Delta_{r}}}dr
\label{Eq:INC}
\end{equation}
the Kerr-de Sitter metric in (\ref{Eq:g}) takes the form:
\begin{equation}
g=-\frac {\Delta_{r}-a^{2} {\Delta_{\vartheta}sin^{2}\vartheta}}{I^{2}\rho^{2}}d^{2}v+\frac {2}{I}dvdr-2 \frac {a}{I}sin^{2}\vartheta
d\overleftarrow{\varphi}dr-2\frac {asin^{2}\vartheta[(r^{2}+a^{2}){\Delta_{\vartheta}-{\Delta_{r}}}]}
{I^{2}\rho^{2}}dvd\overleftarrow{\varphi}+\nonumber
\end{equation}
\begin{equation}
+\frac {{\rho^{2}}} {{\Delta_{\vartheta}}}d^{2}\vartheta+\frac {\Delta_{\vartheta}(r^{2}+a^{2})^{2}-\Delta_{r}a^{2}sin^{2}\vartheta}{I^{2}\rho^{2}}
sin^{2}\vartheta d^{2}\overleftarrow{\varphi}.
\label{Eq:gINC}
\end{equation}
This $g$  is regular  
over points where $\Delta (r)=0$ and by allowing the coordinates $(v,r)$  to run
 over the entire real line, an extension of the Kerr-de Sitter metric is obtained.
In this  $(v, \overleftarrow{\varphi}, r, \vartheta)$ coordinates, the translational $\xi_{t}$ and rotational 
$\xi_{\varphi}$ Killing fields take the form $\xi_{t}=\frac {\partial}{\partial u},\xi_{\varphi}=\frac {\partial}{\partial \overleftarrow{\varphi}}$ 
and the equation $g(\xi_{t},\xi_{t})=0$ shows 
 the existence of non trivial ergospheres.
Their properties depend  
upon the nature of the zeros of $\Delta (r)$ and their significant
 will be discussed elsewhere. Killing Horizons are generated 
  by the Killing field $\hat \xi_{i}=\xi_{t}+\Omega_{i}\xi_{\varphi}$
where as in the case of Kerr, $\Omega_{i}$ are appropriate constants.
These fields become  null precisely over the $r=r_{i}$ hypersurface
and depending upon the values of $(\Lambda, M, a^{2})$ a Kerr-de Sitter spacetime
may contain up to four Killing horizons\footnote{
Promoting these Killing horizons to event horizons
is subtle. For some arguments in that direction see \cite{GibHaw1}.}.
The maximal extension of a Kerr-de Sitter spacetime is obtained by introducing a set of a
outgoing Finkelstein coordinates and joining together these incomplete spacetimes in 
the same manner as for the case a Kerr spacetime. 
Two dimensional conformal diagrams  describing the causal structure of the rotation axis,
can be found for instance in ref
\cite{GibHaw1}, \cite{Mat1}, \cite{Chr1}. In particular in  \cite{Chr1} conformal diagrams for two dimensional sections
of the Kert-de Sitter are analyzed. 
Finally and in view of the comparison between the functions $ \Delta(r)$ and $\hat \Delta(r)$,
the horizon structure  between
a Kerr-de Sitter and a Reissner-Nordstrom-de Sitter spacetime
exhibit similarities. Of course the singularity structure in 
these spacetimes exhibits different features.


\acknowledgments
We thank Majd Abdelqader for discussions related to this work and for sharing with us
his expertise on the invariant representations of the parameters in the Kerr-de Sitter metric.
T.Z. thanks the Department of Physics at Queen's University for hospitality during a sabbatical year. The research of K.L.  was supported in part by a grant from the Natural Sciences and Engineering Research Council of Canada, while the research of T.Z. was supported in part by CONACyT Grant  No. 234571 and by a CIC Grant from the
University of Michoacana, Mexico.

\end{document}